\documentclass[reprint,nofootinbib,amsmath,amssymb,showkeys,aps]{revtex4-1}

\usepackage{graphicx}
\usepackage{bm}
\usepackage{amsmath}
\usepackage{amssymb}
\usepackage{amsfonts}
\usepackage{hyperref}
\usepackage{bbold}
\hypersetup{colorlinks=true,citecolor=blue,linkcolor=blue,urlcolor=blue}
\usepackage{tabularx}
\usepackage{xspace}
\usepackage[normalem]{ulem}
\usepackage{soul}
\usepackage[dvipsnames]{xcolor}

\def\pasp{\ref@jnl{PASP}} % Publications of the ASP
\def\physrep{\ref@jnl{Phys.~Rep.}}   % Physics Reports

\makeatletter
\@ifundefined{textcolor}{}
{%
 \definecolor{BLACK}{gray}{0}
 \definecolor{WHITE}{gray}{1}
 \definecolor{RED}{rgb}{1,0,0}
 \definecolor{GREEN}{rgb}{0,1,0}
 \definecolor{BLUE}{rgb}{0,0,1}
 \definecolor{CYAN}{cmyk}{1,0,0,0}
 \definecolor{MAGENTA}{cmyk}{0,1,0,0}
 \definecolor{YELLOW}{cmyk}{0,0,1,0}
 \definecolor{ao(english)}{rgb}{0.0, 0.5, 0.0}
 \definecolor{applegreen}{rgb}{0.55, 0.71, 0.0}
 \definecolor{brightgreen}{rgb}{0.4, 1.0, 0.0}
}

\def\revtext#1{\textcolor{blue}{#1}}

\def\revtext2#1{\textcolor{applegreen}{#1}}

\begin{document}

\title{Update on Coupled Dark Energy and the $H_0$ tension}

\author{Adri\`a G\'omez-Valent$^1$}\email{gomez-valent@thphys.uni-heidelberg.de}
\author{Valeria Pettorino$^2$}\email{valeria.pettorino@cea.fr}
\author{Luca Amendola$^1$}	\email{l.amendola@thphys.uni-heidelberg.de}

\affiliation{$^1$ Institut f\"{u}r Theoretische Physik, Ruprecht-Karls-Universit\"{a}t Heidelberg, Philosophenweg 16, D-69120 Heidelberg, Germany}
\affiliation{$^2$ AIM, CEA, CNRS, Universit{\'e} Paris-Saclay, Universit{\'e} Paris Diderot, 
             Sorbonne Paris Cit{\'e}, F-91191 Gif-sur-Yvette, France}

%\date{\today}

\begin{abstract}
In this work we provide updated constraints on coupled dark energy (CDE) cosmology with Peebles-Ratra (PR) potential and constant coupling strength $\beta$. This modified gravity scenario introduces a fifth force between dark matter particles, mediated by a scalar field that plays the role of dark energy. The mass of the dark matter particles does not remain constant, but changes with time as a function of the scalar field. Here we focus on the phenomenological behavior of the model, and assess its ability to describe updated cosmological data sets that include the {\it Planck} 2018 cosmic microwave background (CMB) temperature, polarization and lensing, baryon acoustic oscillations, the Pantheon compilation of supernovae of Type Ia, data on $H(z)$ from cosmic chronometers, and redshift-space distortions. We also study which is the impact of the local measurement of $H_0$ from SH0ES and the strong-lensing time delay data from the H0LICOW collaboration on the parameter that controls the strength of the interaction in the dark sector. We find a peak corresponding to a coupling $\beta > 0$ and to a potential parameter $\alpha > 0$, more or less evident depending on the data set combination. We show separately the impact of each data set and remark that it is especially CMB lensing the one data set that shifts the peak the most towards $\Lambda$CDM. When a model selection criterion based on the full Bayesian evidence is applied, however, $\Lambda$CDM is still preferred in all cases, due to the additional parameters introduced in the CDE model.
\end{abstract}

%\pacs{}

\maketitle

%%%%%%%%%%%%%%%%%%%%%%%%%%%%%%%%%%%%%%%%%%%%%%%%%%%%%%%%%%%%%%%%%
%%%%%%%%%%%%%%%%%%%%%%%%%%%%%%%%%%%%%%%%%%%%%%%%%%%%%%%%%%%%%%%%%
%%%%%%%%%%%%%%%%%%%%%%%%%%%%%%%%%%%%%%%%%%%%%%%%%%%%%%%%%%%%%%%%%

\section{Introduction}\label{intro}

Important observational hints in favor of the positive acceleration of the Universe appeared already more than twenty years ago, thanks to the detection of standardizable high-redshift supernovae of Type Ia (SNIa) and the measurement of their light-curves and redshifts \cite{Riess:1998cb,Perlmutter:1998np}. Since then, many other probes have contributed to increase the evidence in favor of the late-time accelerated phase. They range e.g. from the detection of the baryon acoustic peak in the two-point correlation function of matter density fluctuations \cite{Cole:2005sx,Eisenstein:2005su} to the very accurate measurement of the cosmic microwave background (CMB) temperature anisotropies by WMAP \cite{Hinshaw:2012aka} and {\it Planck} \cite{Ade:2013sjv,Ade:2015xua,Aghanim:2018eyx}. At the phenomenological level, the easiest explanation for such acceleration is given by the presence of a very tiny cosmological constant in Einstein's field equations, with an associated energy density which is orders of magnitude lower than the quantum field theoretical estimates made for the vacuum energy density. Protecting such low value from radiative corrections is extremely difficult and constitutes the core of the so-called ``old'' cosmological constant problem, cf. e.g. \cite{Weinberg:1988cp,Martin:2012bt,Sola:2013gha}. In addition, explaining why the current value of this energy density is of the same order of magnitude as the matter energy density, the so-called  ``coincidence problem'', is considered by part of the cosmological community as another problem that needs to be addressed.  The cosmological constant is a pivotal ingredient of the standard cosmological model, also known as $\Lambda$CDM or concordance model (cf. e.g. the reviews \cite{Peebles:2002gy,Padmanabhan:2002ji}), which can explain most of the cosmological observations with high proficiency. Nevertheless, the aforementioned theoretical conundrums, together with few persistent tensions in some relevant parameters of the model as the Hubble parameter $H_0$ \cite{Aghanim:2018eyx,Riess:2019cxk} and the root-mean-square ({\it rms}) of mass fluctuations at scales of $8h^{-1}$ Mpc \cite{Macaulay:2013swa}, $\sigma_8$ (or $S_8=\sigma_8(\Omega_m^{(0)}/0.3)^{0.5}$ \footnote{The superscripts $(0)$ will denote from now on quantities evaluated at present, i.e. at $a=1$.} \cite{Hildebrandt:2018yau}), with $h$ being the reduced Hubble parameter , motivate theoretical cosmologists to look for alternative scenarios in which these problems can be solved or, at least, alleviated,  see \cite{AmendolaTsujikawaBook,Joyce:2014kja} and references therein. Wherever the solution comes from, i.e. a departure from General Relativity or some sort of new field  describing dark energy (DE), it must mimic very well the behavior of a cosmological constant at low redshifts, meaning that the corresponding effective equation of state (EoS) parameter must be very close to -1, and that the new component must not be able to cluster efficiently at low scales.

%%%%%%%%%%%%%%%%%%%%%%%%%%%%%%%%%%%%%%%%%%%%%%%%%%%%%%%%%%%%%%%%%
%%%%%%%%%%%%%% FIGURE 1 %%%%%%%%%%%%%%%%%%%%%%%%%%%
%%%%%%%%%%%%%%%%%%%%%%%
%%%%
\begin{figure*}
\begin{center}
\includegraphics[width=7in, height=2.4in]{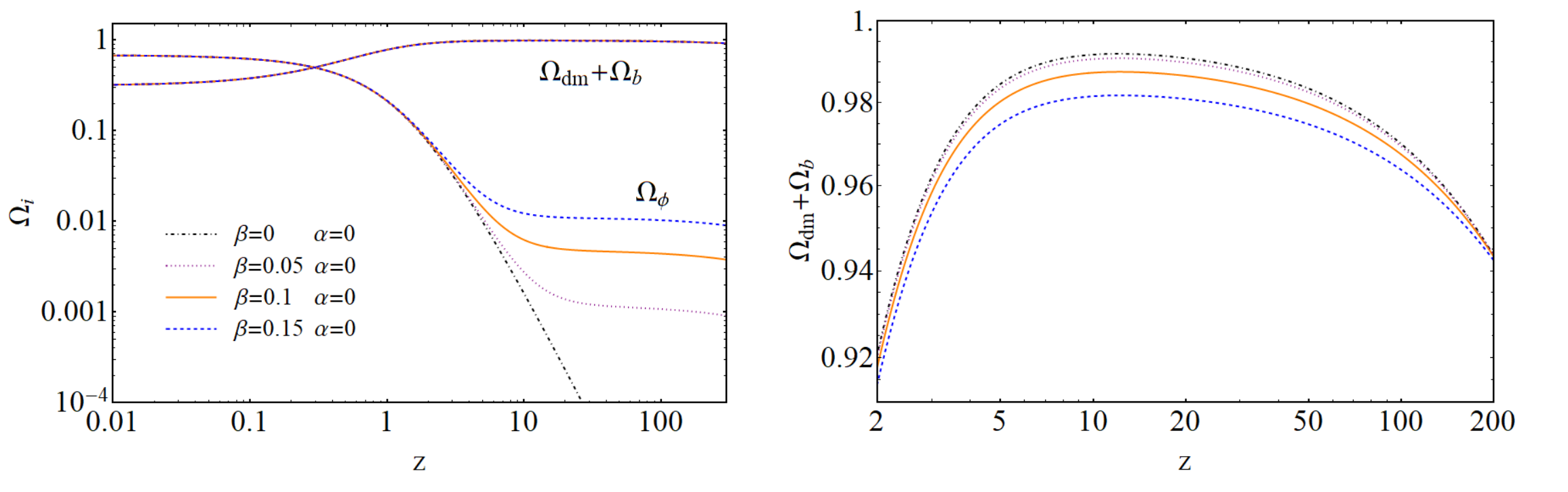}
\caption{\scriptsize {{\it Left plot:} Normalized densities $\Omega_{dm}(z)+\Omega_{b}(z)$ and $\Omega_\phi(z)$ for four alternative values of $\beta$ and considering a constant potential. The other parameters (including the current energy densities) have been set to the best-fit $\Lambda$CDM values from the TTTEEE+lowE {\it Planck} 2018 analysis \cite{Aghanim:2018eyx}. {\it Right plot:} Here we zoom in the range $z=[2,200]$ of the $\Omega_{dm}+\Omega_b$ curves in order to better visualize their evolution during the matter-dominated epoch, when the system is near the $\phi$MDE fixed point. See the text for details.}}
\label{fig:Omegas}                                                      
\end{center}
\end{figure*}
%%%%%%
%%%%%%%%%%%%%%%%%%%%%%%%%%%%%%%%%%%%%%%%%%%%%%%%%%%%%%%%%%%%%%%%%
%%%%%%%%%%%%%%%%%%%%%%%%%%%%%%%%%%%%%%%%%%%%%%%%%%%%%%%%%%%%%%%%%

In this paper we consider a scenario in which dark matter (DM) particles interact via a force mediated by a scalar field, which in turn drives cosmic acceleration. This scenario is referred to as \textit{coupled dark energy} (CDE). It was originally proposed as a means of alleviating the coincidence problem \cite{Wetterich:1994bg,Amendola:1999er}, considering not only a potential energy density for quintessence to generate its dynamics, but also allowing an interaction with other sectors of the theory. These interactions extended the original quintessence models \cite{Peccei:1987mm,Wetterich:1987fm,Peebles:1987ek,Ratra:1987rm}. They cannot be ruled out {\it a priori} and, hence, they must be duly constrained by experiments and observations.

Some works already set constraints  on this model, but using   older cosmological data sets, for instance CMB data from the WMAP satellite and the South Pole Telescope \cite{Pettorino:2012ts}, or considering past (2013, 2015) releases of {\it Planck} CMB data in combination with other data sets, as e.g. from baryon acoustic oscillations (BAO) and SNIa, \cite{Pettorino:2013oxa,Ade:2015rim}. Intriguingly, these works detected a likelihood peak at a non-vanishing value of the coupling constant. One of the main goals of this paper is then to critically revisit and update these results in the light of the recent strengthening of the $H_0$ tension and of the rich amount of currently available data at our disposal, in particular the {\it Planck} 2018 CMB temperature, polarization and lensing data, but also other new cosmological data, for instance Refs. \cite{Wong:2019kwg,Gil-Marin:2018cgo}. For constraints on other models with DM-DE interactions see e.g.  \cite{Xia:2013nua,Pourtsidou:2016ico,vandeBruck:2016hpz,vandeBruck:2017idm,Li:2014cee,Li:2015vla,DiValentino:2017iww,Abdalla:2014cla,Costa:2016tpb,Sola:2017znb,Sola:2017jbl,Martinelli:2019dau,Agrawal:2019dlm,Pan:2020zza}, and when the interaction is motivated in the context of the running vacuum models \cite{Sola:2017jbl,Sola:2017znb,Sola:2016ecz,Gomez-Valent:2018nib,Tsiapi:2018she}.

%%%%%%%%%%%%%%%%%%%%%%%%%%%%%%%%%%%%%%%%%%%%%%%%%%%%%%%%%%%%%%%%%
%%%%%%%%%%%%%%%%%%%%%%%%%%%%%%%%%%%%%%%%%%%%%%%%%%%%%%%%%%%%%%%%%
%%%%%%%%%%%%%%%%%%%%%%%%%%%%%%%%%%%%%%%%%%%%%%%%%%%%%%%%%%%%%%%%%

\section{Coupled dark energy}\label{sect:CDEmodel}

We consider a CDE scenario, as studied in \cite{Amendola:1999er,Amendola:2003wa,Pettorino:2008ez}, to which we refer for a detailed description. We here briefly recall the main equations. This CDE model is formulated in the so-called Einstein or observational frame \cite{Amendola:2019xqj}. Apart from the Standard Model of Particle Physics and a potential extension accounting for the origin of the neutrino masses, we consider a dark sector described by the following Lagrangian density:
\begin{equation}\label{eq:DarkLagrangian}
\mathcal{L}_{\rm dark} = -\partial_\mu\phi\partial^\mu\phi-V(\phi)-m(\phi)\bar{\psi}\psi+\mathcal{L}_{\rm kin}[\psi]\,,
\end{equation}
where $\phi$ is the scalar field that plays the role of DE, with potential $V(\phi)$, and $\psi$ is the DM field, considered here to be of fermionic nature, just for illustrative purposes. The DM particles interact with the DE due to the $\phi$-dependent mass term appearing in \eqref{eq:DarkLagrangian}. Such interaction introduces a fifth force that alters the trajectory in space-time of the DM with respect to the one found in the uncoupled case. Depending on the strength of the force, this model can be force-accelerated, as opposed to fluid-accelerated, adopting the terminology of \cite{Amendola:2019xqj}. As we do not couple $\phi$ to the standard model sector we avoid the stringent local (solar system) constraints on the violation of the weak equivalence principle \cite{Will:2005va}, and also on screened fifth forces that couple $\phi$ to non-dark matter, e.g. from Casimir experiments \cite{Elder:2019yyp}, precision measurements of the electron magnetic moment \cite{Brax:2018zfb}, or measurements of the E\"{o}tv\"{o}s parameter \cite{Berge:2017ovy}. They have no impact on the CDE model under study. 

%%%%%%%%%%%%%%%%%%%%%%%%%%%%%%%%%%%%%%%%%%%%%%%%%%%%%%%%%%%%%%%%%
%%%%%%%%%%%%%% FIGURE 2 %%%%%%%%%%%%%%%%%%%%%%%%%%%
%%%%%%%%%%%%%%%%%%%%%%%
%%%%
\begin{figure*}
\begin{center}
\includegraphics[width=7in, height=2.4in]{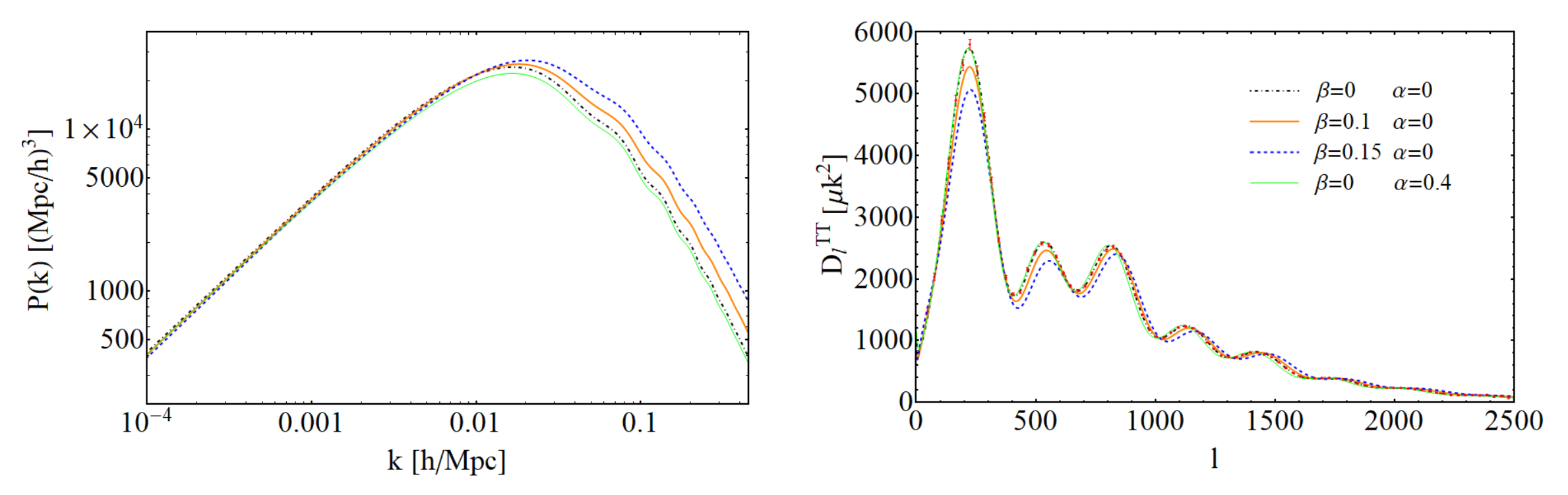}
\caption{\scriptsize {Theoretical curves of the current matter power spectrum (left plot) and CMB temperature anisotropies (right plot) for the $\Lambda$CDM, two CDE models with $\beta=0.1,\,0.15$ and flat potential, and also for the uncoupled Peebles-Ratra model with $\alpha=0.4$. We set the other parameters as in \autoref{fig:Omegas}. In the right plot we also include the observational data from \cite{Aghanim:2018eyx} (in red). These figures show: (i) the enhancement of the growth of matter perturbations caused by $\beta>0$, and the opposite effect produced by $\alpha>0$; and (ii) the shift to larger multipoles and the amplitude suppression of the acoustic peaks induced by increasing values of $\beta$. See the text for further details.}} 
\label{fig:ClsFig}                                          
\end{center}
\end{figure*}
%%%%%%
%%%%%%%%%%%%%%%%%%%%%%%%%%%%%%%%%%%%%%%%%%%%%%%%%%%%%%%%%%%%%%%%%
%%%%%%%%%%%%%%%%%%%%%%%%%%%%%%%%%%%%%%%%%%%%%%%%%%%%%%%%%%%%%%%%%

The variation of the total action with respect to the metric leads as usual to Einstein's equations, and the covariant energy of the joint system DM-DE is conserved. Hence, $\nabla^\mu T^{\phi}_{\mu\nu}= +Q_\nu$ and $\nabla^\mu T^{dm}_{\mu\nu}=-Q_\nu$, with $Q_\nu$ defined as 
\begin{equation}\label{eq:source}
Q_\nu = \beta \kappa T^{dm}\nabla_\nu\phi\,,
\end{equation}
where $\kappa=\sqrt{8\pi G}$, $T^{dm}$ is the trace of the DM energy-momentum tensor, and $\beta$ controls the strength of the interaction and is in general a function of $\phi$. If set to zero, we recover the equations of uncoupled quintessence. In this work we consider $\beta$ to be a positive constant.

We assume that the Universe is spatially flat, as supported by CMB information from {\it Planck} 2018 when combined with BAO \cite{Aghanim:2018eyx} and/or SNIa \cite{Efstathiou:2020wem}, with the curvature parameter $\Omega^{(0)}_K$ constrained to be lower than $\sim 2\%$ at $68\%$ c.l. in $\Lambda$CDM. Thus, we can make use of the Friedmann-Lema\^itre-Robertson-Walker metric, which at the background level reads $ds^2=a^2(\tau)\left[-d\tau^2+\delta_{ij}dx^i dx^j\right]$, with $a$ being the scale factor, $\tau$ the conformal time, and $x^i$ for $i=1,2,3$ the spatial comoving coordinates. In addition, we treat DM as a pressureless perfect fluid, so the conservation equations for DE and DM can be written, respectively,
\begin{equation}\label{eq:KGeq}
\beta\kappa a^2\rho_{dm} = \phi^{\prime\prime}+2\mathcal{H}\phi^\prime+a^2\frac{\partial V}{\partial\phi}\,,
\end{equation}
\begin{equation}\label{eq:consDM}
\rho^\prime_{dm}+3\mathcal{H}\rho_{dm}=-\beta\kappa\rho_{dm}\phi^\prime\,,
\end{equation}
with $\rho_{dm}$ the DM energy density, $\mathcal{H}=a^\prime/a$, and the primes denoting derivatives {\it w.r.t.} the conformal time. All the functions entering these equations are background quantities. If we assume the conservation of the number density of DM particles then their mass evolves as $m(\phi)=m^{(0)}e^{\beta\kappa(\phi^{(0)}-\phi)}$.

A feature of the model is that for $\beta^2<3/2$ it has an unstable (saddle) fixed point at $(\Omega_{dm},\Omega_{\phi})=(1-2\beta^2/3,2\beta^2/3)$, where $\Omega_i=\rho_i/\rho_c$, with $\rho_c$ the critical energy density. This fixed point (dubbed $\phi$MDE in \cite{Amendola:1999er}) cannot be reached exactly, since there is also a non-null fraction of baryons, but the system can be quite close to it, since the DM energy density is much larger than the baryonic one (cf. \autoref{fig:Omegas}). During this phase the effective EoS parameter, i.e. the ratio of the total pressure and the critical energy density in the Universe, is given by $w_{\rm eff}=\Omega_{\phi}$, and hence the deceleration parameter reads $q=\frac{1}{2}(1+3\,w_{\rm eff})=\frac{1}{2}+\beta^2$. Thus, the coupling between DM and DE makes the Universe more decelerated with respect to the uncoupled quintessence case during the matter-dominated epoch (MDE). This fact together with the fifth force that enters now as a new source term in the Poisson equation help matter inhomogeneities to grow faster for larger values of $\beta$. We also remark that for fixed values of the present energy densities, matter becomes dominant over radiation earlier in time when $\beta>0$, with respect to the uncoupled case. In the CDE scenario, the equation for the DM density contrast $\delta_{dm}=\delta\rho_{dm}/\rho_{dm}$ at deep subhorizon scales ($k\gg \mathcal{H}$) and when non-linear processes are unimportant, reads,
\begin{eqnarray}\label{eq:DMcontrast}
\delta_{dm}^{\prime\prime}+(\mathcal{H}-&&\beta\kappa\phi^\prime)\delta_{dm}^\prime\phantom{XXXX}\nonumber\\
&&-4\pi Ga^2[\rho_b\delta_b+\rho_{dm}\delta_{dm}(1+2\beta^2)]=0\,.\phantom{XX}
\end{eqnarray}
If we neglect the contribution of baryons, $\delta_m(a)\sim a^{1+2\beta^2}$. Hence, larger values of $\beta$ enhance the matter power spectrum (see the left plot of  \autoref{fig:ClsFig}) and leave an imprint on the CMB temperature anisotropies. First, the integrated Sachs-Wolfe effect \cite{Sachs:1967er}  is enhanced during the MDE earlier than in the uncoupled scenario, in which such effect is only relevant after matter-domination; second, the coupling affects lensing of CMB by large scale structure; the interaction also shifts the position of the acoustic peaks to larger multipoles due to the decrease of the sound horizon at the baryon-drag epoch, which is caused by the increase of the mass of the DM particles (this latter effect is however typically very small and subdominant). Finally, the amplitude is suppressed, because of the decrease of $\rho_b/\rho_{dm}$ at recombination. These two effects explain why the coupling strength is degenerate with the Hubble parameter today \cite{Pettorino:2013oxa}, whose value is related to the position and overall amplitude of the first peak. These and other aspects of the structure formation were already discussed in \cite{Pettorino:2008ez, Amendola:2011ie, Baldi:2008ay, Baldi:2010td}. See therein for further details, and also the plots in \autoref{fig:ClsFig}.

The quintessence potential only rules the dynamics of $\phi$ in the late-time universe, after the MDE, when the interaction term appearing in the {\it l.h.s.} of \eqref{eq:KGeq} becomes subdominant. It helps to slow down structure formation processes {\it w.r.t.} the flat-potential scenario (for a fixed value of the current DE density). Hence, it can compensate in lesser or greater extent (depending on its steepness) the enhancement of power generated by the fifth force during the MDE (cf. the left plot of \autoref{fig:ClsFig} and its caption). 

We employ the Peebles-Ratra (PR) potential \cite{Peebles:1987ek,Ratra:1987rm}, 
\begin{equation}\label{eq:PeeblesRatra}
V(\phi) = V_0\phi^{-\alpha}\,,
\end{equation}
with $V_0$ and $\alpha>0$ being constants, and the former having dimensions of mass$^{4+\alpha}$ in natural units, since $\phi$ has dimensions of mass. We want to update the constraints on the parameters of the CDE model with PR potential that were obtained in some past works using older CMB data, from WMAP and/or past releases of {\it Planck} (cf. \cite{Amendola:2003eq,Pettorino:2012ts,Pettorino:2013oxa,Ade:2015rim}), so it is natural to stick to \eqref{eq:PeeblesRatra} here. Also because it has proved to be capable of improving the description of some cosmological data sets with respect to the $\Lambda$CDM model in the non-interactive case \cite{Sola:2016hnq,Ooba:2018dzf,Sola:2018sjf}. 

The CDE model we are considering (i.e. CDE with PR potential) has three nested models, namely the $\Lambda$CDM, the PR model, and the CDE model with flat potential. They are obtained from the full CDE model with \eqref{eq:PeeblesRatra} in the limits $(\alpha,\beta)\to (0,0)$, $\beta\to 0$ and $\alpha\to 0$, respectively. We also provide constraints on these scenarios in appendix B. 

For recent studies on CDE with an exponential potential, see \cite{Xia:2013nua,vandeBruck:2016hpz,vandeBruck:2017idm,Agrawal:2019dlm}. We report fitting results for this model too, in appendix C.

%%%%%%%%%%%%%%%%%%%%%%%%%%%%%%%%%%%%%%%%%%%%%%%%%%%%%%%%%%%%%%%%%
%%%%%%%%%%%%%%%%%%%%%%%%%%%%%%%%%%%%%%%%%%%%%%%%%%%%%%%%%%%%%%%%%
%%%%%%%%%%%%%%%%%%%%%%%%%%%%%%%%%%%%%%%%%%%%%%%%%%%%%%%%%%%%%%%%%

\section{Methodology}\label{sect:methodology}
We have implemented the CDE model described in \autoref{sect:CDEmodel} in our own modified version of the Einstein-Boltzmann system solver \texttt{CLASS}\footnote{http://lesgourg.github.io/class\_public/class.html} \cite{Blas:2011rf}, which allows us to solve the background and linear perturbations equations and produce the theoretical quantities of interest, as e.g. the matter power spectrum, the CMB anisotropies, the cosmological distances, etc. This implementation has been compared and validated with the interacting dark energy anisotropy (IDEA) code, used in \cite{Pettorino:2012ts,Pettorino:2013ia,Amendola:2007yx,Ade:2015rim}. The Bayesian exploration of the parameter space of the model in the light of the various data sets described in \autoref{sect:data} has been carried out with the Monte Carlo sampler \texttt{Montepython}\footnote{http://baudren.github.io/montepython.html} \cite{Audren:2012wb}. Our code lets us skip the shooting method that is employed in IDEA to match the initial conditions with the current values of the cosmological energy densities, and this allows us to improve the computational efficiency of our Markov chain Monte Carlo (MCMC) analyses, cf. appendix A for details. We have also used the \texttt{Python} package \texttt{GetDist}\footnote{https://getdist.readthedocs.io/en/latest/} \cite{Lewis:2019xzd} to process the chains and obtain the mean values and uncertainties of the parameters reported in Tables \ref{tab:tab1}-\ref{tab:tab3}, as well as the contours of Figs. \ref{fig:betaCLs}-\ref{fig:nolensVSlens_reduced}. Finally, we have computed the full Bayesian evidences for all the models and under the various data sets, by processing the corresponding Markov chains with the code \texttt{MCEvidence}\footnote{https://github.com/yabebalFantaye/MCEvidence} \cite{Heavens:2017afc}. This has allowed us to carry out a rigorous model comparison analysis, which we present in \autoref{sect:results}.

%%%%%%%%%%%%%%%%%%%%%%%%%%%%%%%%%%%%%%%%%%%%%%%%%%%%%%%%%%%%%%%%%
%%%%%%%%%%%%%%%%%%%%%%%%%%%%%%%%%%%%%%%%%%%%%%%%%%%%%%%%%%%%%%%%%

\section{Data}\label{sect:data}
Since the last fitting analysis of the CDE model with PR potential, in \cite{Ade:2015rim}, new and more precise data have appeared in the literature. In this paper we perform an exhaustive update of the data sets with respect to those used in \cite{Ade:2015rim}. The most important changes are: (i) here we make use of the {\it Planck} 2018 CMB data \cite{Aghanim:2018eyx} instead of the 2015 release \cite{Ade:2015xua}; (ii) we fully update our BAO and redshift-space distortions (RSD) data sets, using now e.g. the data of the last release of the Baryon Oscillation Spectroscopic Survey\footnote{http://www.sdss3.org/surveys/boss.php} (BOSS); (iii) we substitute the SNIa sample from the Joint-Light-curve Analysis (JLA) \cite{Betoule:2014frx} by the Pantheon+MCT compilation \cite{Scolnic:2017caz,Riess:2017lxs}, which contains the former and includes 323 additional SNIa; (iv) we study the impact of the data on $H(z)$ obtained from cosmic chronometers (CCH); (v) instead of using the prior on $H_0$ from \cite{Efstathiou:2013via}, $H_0=(70.6\pm 3.3)$ km/s/Mpc, we use the measurement by the SH0ES collaboration reported in \cite{Riess:2019cxk} (see \autoref{sect:H0} and comments therein); and (vi) we also study the effect that the inclusion of the strong-lensing time delay distances measured by H0LICOW has on our constraints. We use, therefore, a much richer data set here than the one employed in \cite{Ade:2015rim}.

Our data set is very similar to the one used by the {\it Planck} collaboration in their 2018 analysis of the $\Lambda$CDM and minimal extensions of it \cite{Aghanim:2018eyx}. There are some differences, though, e.g. we analyze here the effect of cosmic chronometers and the H0LICOW data, something that was not done there. We refer the reader to \autoref{sect:descData} and reference \cite{Aghanim:2018eyx} for details. 

%%%%%%%%%%%%%%%%%%%%%%%%%%%%%%%%%%%%%%%%%%%%%%%%%%%%%%%%%%%%%%%%%
%%%%%%%%%%%%%%%%%%%%%%%%%%%%%%%%%%%%%%%%%%%%%%%%%%%%%%%%%%%%%%%%%

\subsection{Description of the individual data sets}\label{sect:descData}
Here we list and describe the individual data sets that we employ in this work to constrain the CDE model that we have presented in \autoref{sect:CDEmodel}, and its nested models. We will study their impact by considering different data set combinations, as explained in \autoref{sect:CombiData}.

%%%%%%%%%%%%%%%%%%%%%%%%%%%%%%%%%%%%%%%%%%%%%%%%%%%%%%%%%%%%%%%%%

\subsubsection{Cosmic microwave background}\label{sect:CMB}
We derive all the main results of this paper making use of the full TTTEEE+lowE CMB likelihood from \textit{Planck} 2018 \cite{Aghanim:2018eyx}, which includes the data on the CMB temperature (TT) and polarization (EE) anisotropies, and their cross-correlations (TE) at both low and high multipoles. We also study what is the impact of also including the CMB lensing likelihood \cite{Aghanim:2018oex}. Temperature and polarization spectra are already lensed, however the CMB lensing likelihood includes on top of lensed spectra also the 4-point correlation function.  Lensing peak sensitivity is to lenses
at $z\approx 2$, half-way to the last-scattering surface, with deflection effects at redshifts which are relevant for dark energy models such as CDE. It has in particular been shown in \cite{Ade:2015rim} that CMB lensing pushes constraints towards $\Lambda$CDM. As stated in \cite{Aghanim:2018eyx}, we note that the lensing likelihood assumes a fiducial $\Lambda$CDM model, with linear corrections to the fiducial model accounted for self-consistently. According to \cite{Aghanim:2018eyx}
this procedure is unbiased, at least up to when the lensing spectrum differs from the fiducial spectrum by as much as 20$\%$, estimated to be larger than differences allowed by the CMB lensing data. While further independent validation of such tests would be interesting for future analyses on modified gravity, we find it important to comment on results with/without CMB lensing inclusion for the purpose of testing non-minimal extensions of $\Lambda$CDM, such as CDE.

%%%%%%%%%%%%%%%%%%%%%%%%%%%%%%%%%%%%%%%%%%%%%%%%%%%%%%%%%%%%%%%%%

\subsubsection{Baryon acoustic oscillations}\label{sect:BAO}
Baryon acoustic oscillations are a direct consequence of the strong coupling between photons and baryons in the pre-recombination epoch. After the decoupling of photons, the overdensities in the baryon fluid evolved and attracted more matter, leaving an imprint in the two-point correlation function of matter fluctuations with a characteristic scale of around $147$ Mpc that can be used as a standard ruler and to constrain cosmological models. It was firstly measured by \cite{Cole:2005sx,Eisenstein:2005su} using the galaxy power spectrum. Since then, several galaxy surveys have been able to provide precise data on BAO, either in terms of the dilation scale $D_V$,
\begin{equation}
\frac{D_V(z)}{r_d}=\frac{1}{r_d}\left[D_M^2(z)\frac{cz}{H(z)}\right]^{1/3}\,,
\end{equation}
with $D_M=(1+z)D_{A}(z)$ being the comoving angular diameter distance and $r_d$ the sound horizon at the baryon drag epoch, or even by splitting (when possible) the transverse and line-of-sight BAO information and hence being able to provide data on $D_{A}(z)/r_d$ and $H(z)r_d$ separately, with some degree of correlation. The surveys provide the values of the measurements at some effective redshift(s). We employ the following BAO data points:

\begin{itemize}

\item $D_V/r_d$ at $z=0.122$ provided in \cite{Carter:2018vce}, which combines the dilation scales previously reported by the 6dF Galaxy Survey\footnote{http://www.6dfgs.net/} (6dFGS) \cite{Beutler:2011hx} at $z=0.106$ and the one obtained from the Sloan Digital Sky Survey\footnote{https://www.sdss.org/} (SDSS) Main Galaxy Sample at $z=0.15$ \cite{Ross:2014qpa}.

\item The anisotropic BAO data measured by BOSS using the LOWZ ($z=0.32$) and CMASS ($z=0.57$) galaxy samples \cite{Gil-Marin:2016wya}.

\item The dilation scale measurements by WiggleZ\footnote{http://wigglez.swin.edu.au/site/} at $z=0.44,0.60,0.73$ \cite{Kazin:2014qga}. The galaxies contained in the WiggleZ catalog are located in a patch of the sky that partially overlaps with those present in the CMASS sample by BOSS. Nevertheless, the two surveys are independent, work under different seeing conditions, instrumental noise, etc. and target different types of galaxies. The correlation between the CMASS and WiggleZ data has been quantified in \cite{Beutler:2015tla}, were the authors estimated the correlation coefficient to be $\lesssim 2\%$. This justifies the inclusion of the WiggleZ data in our analysis, although their statistical weight is much lower than those from BOSS and in practice their use does not have any important impact on our results. 

\item $D_A/r_d$ at $z=0.81$ measured by the Dark Energy Survey (DES)\footnote{https://www.darkenergysurvey.org/es/} \cite{Abbott:2017wcz}.

\item  The anisotropic BAO data from the extended BOSS Data Release (DR) 14 quasar sample at $z=1.19,1.50,1.83$ \cite{Gil-Marin:2018cgo}.

\item The combined measurement of the anisotropic BAO information obtained from the Ly$\alpha$-quasar cross and auto-correlation of eBOSS DR14 \cite{Blomqvist:2019rah,Agathe:2019vsu}, at $z=2.34$. 

\end{itemize}

%%%%%%%%%%%%%%%%%%%%%%%%%%%%%%%%%%%%%%%%%%%%%%%%%%%%%%%%%%%%%%%%%

\subsubsection{Supernovae of Type Ia}\label{sect:SNIa}
We consider $6$ effective points on the Hubble rate, i.e. $E(z)\equiv H(z)/H_0$, and the associated covariance matrix. They compress the information of 1048 SNIa contained in the Pantheon compilation \cite{Scolnic:2017caz} and the 15 SNIa at $z>1$ from the Hubble Space Telescope Multi-Cycle Treasury programs \cite{Riess:2017lxs}. The compression effectiveness of the information contained in such SNIa samples is extremely good, as it is explicitly shown in \cite{Riess:2017lxs}. See, e.g. Fig. 3 of that reference and the corresponding explanations in the main text.

%%%%%%%%%%%%%%%%%%%%%%%%%%%%%%%%%%%%%%%%%%%%%%%%%%%%%%%%%%%%%%%%%

\subsubsection{Cosmic chronometers}\label{sect:CCH}
Spectroscopic dating techniques of passively–evolving galaxies, i.e. galaxies with old stellar populations and low star formation rates, have become a good tool to obtain observational values of the Hubble function at redshifts $z\lesssim 2$ \cite{Jimenez:2001gg}. These measurements do not rely on any particular cosmological model, although are subject to other sources of systematic uncertainties, as to the ones associated to the modeling of stellar ages, see e.g. \cite{Moresco:2012jh,Moresco:2016mzx}, which is carried out through the so-called stellar population synthesis (SPS) techniques, and also to a possible contamination due to the presence of young stellar components in such quiescent galaxies \cite{Lopez-Corredoira:2017zfl,Lopez-Corredoira:2018tmn,Moresco:2018xdr}. Given a pair of ensembles of passively-evolving galaxies at two diﬀerent redshifts it is possible to infer $dz/dt$ from observations under the assumption of a concrete SPS model and compute $H(z) = -(1 + z)^{-1}dz/dt$. Thus, cosmic chronometers allow us to obtain the value of the Hubble function at different redshifts, contrary to other probes which do not directly measure $H(z)$, but integrated quantities as e.g. luminosity distances.

In this work we use the 31 data points on $H(z)$ from CCH provided in \cite{Jimenez:2003iv,Simon:2004tf,Stern:2009ep,Moresco:2012jh,Zhang:2012mp,Moresco:2015cya,Moresco:2016mzx,Ratsimbazafy:2017vga}. More concretely, we make use of the {\it processed} sample provided in Table 2 of \cite{Gomez-Valent:2018gvm}, which is more conservative, since it introduces corrections accounting for the systematic errors mentioned above.

Several authors have employed these data to reconstruct the expansion history of the Universe using Gaussian Processes and/or the so-called Weighted Function Regression method \cite{Yu:2017iju,Gomez-Valent:2018hwc,Haridasu:2018gqm}. These approaches do not rely on a particular cosmological model. They find extrapolated values of the Hubble parameter that are closer to the best-fit $\Lambda$CDM value reported by {\it Planck} \cite{Aghanim:2018eyx}, around $H_0\sim (67.5-69.5)$ km/s/Mpc, but still compatible at $\sim 1\sigma$ c.l. with the local determination obtained with the distance ladder technique \cite{Riess:2018uxu,Riess:2019cxk}. When BAO data and/or the SNIa from the Pantheon compilation are also incorporated in the analyses together with the CCH, the tension between the local measurement and the one inferred from the reconstruction arises again, but only at a small $\sim 2\sigma$ c.l. \cite{Yu:2017iju,Gomez-Valent:2018hwc,Haridasu:2018gqm}.

%%%%%%%%%%%%%%%%%%%%%%%%%%%%%%%%%%%%%%%%%%%%%%%%%%%%%%%%%%%%%%%%%

\subsubsection{Redshift-space distortions}\label{sect:RSD}

Measurements of the peculiar velocities of galaxies can be obtained from observations of their anisotropic clustering in redshift space. They allow galaxy redshift surveys to obtain constraints on the product of the growth rate of structure, $f(z)=\frac{d\ln \delta_m(a)}{d\ln a}$, and the {\it rms} of mass fluctuations at scales of $8h^{-1}$ Mpc, $\sigma_8(z)$. Much of the statistical signal comes, though, from scales where nonlinear effects and galaxy bias are signiﬁcant and they must be accurately modeled. The modeling techniques have been improved in the last years, making data on RSD to be a reliable tool to constrain cosmological models. These are the measurements that we include in our RSD data set:

\begin{itemize}

\item  The data point at $z=0.03$ obtained upon combining the density and velocity ﬁelds measured by the 2MASS Tully-Fisher (2MTF) and 6dFGS peculiar-velocity surveys \cite{Qin:2019axr}.

\item The point reported by SDSS DR7 at $z=0.1$ \cite{Shi:2017qpr}.

\item The two data points provided by the Galaxy and Mass Assembly survey (GAMA) at $z=0.18$ \cite{Simpson:2015yfa} and $z=0.38$ \cite{Blake:2013nif}.

\item The four points at $z=0.22,0.41,0.60,0.78$ measured by WiggleZ \cite{Blake:2011rj}.

\item  The RSD measurements by BOSS from the power spectrum and bispectrum of the DR12 galaxies contained in the LOWZ ($z=0.32$) and CMASS ($0.57$) samples \cite{Gil-Marin:2016wya}.

\item The two points at $z=0.60,0.86$ reported by the VIMOS Public Extragalactic Redshift Survey (VIPERS) \cite{Mohammad:2018mdy}.

\item The point at $z=0.77$ by VIMOS VLT Deep Survey (VVDS) \cite{Guzzo:2008ac,Song:2008qt}. 

\item The measurement by eBOSS DR14 at $z=1.19$, $1.50$, $1.83$ \cite{Gil-Marin:2018cgo}.

\item The Subaru FMOS galaxy redshift survey (FastSound) measurement at $z=1.36$ \cite{Okumura:2015lvp}.

\end{itemize}

%
%%%%%%%%%%%%%%%%%%%%%%%%%%%%%%%%%%%%%%%%%%%%%%%%%%%%%%%%%%%%%%%%%%%%%%%%%%%%%%
%
\begin{table*}[t]
\begin{center}
\resizebox{17.8cm}{!}{
\begin{tabular}{| c | c |c | c | c| c| c|c|}
\multicolumn{1}{c}{Parameter} &  \multicolumn{1}{c}{P18} & \multicolumn{1}{c}{P18+BSC}   & \multicolumn{1}{c}{P18+SH0ES+H0LICOW} & \multicolumn{1}{c}{P18+BSC+RSD} & \multicolumn{1}{c}{P18lens+BSC+RSD} &  \multicolumn{1}{c}{P18+BSC+SH0ES+H0LICOW} & \multicolumn{1}{c}{P18lens+SH0ES+H0LICOW}
\\\hline
$\Omega^{(0)}_{dm}h^2$ & $0.1207^{+0.0014}_{-0.0013}$ & $0.1192\pm 0.0008$  & $0.1172^{+0.0012}_{-0.0014}$ & $0.1187\pm 0.0008$ & $0.1191\pm 0.0007$ & $0.1185\pm 0.0008$ & $0.1182^{+0.0011}_{-0.0010}$
\\\hline
$\Omega^{(0)}_{b}h^2$ & $0.02237\pm 0.00015$ & $0.02242^{+0.00010}_{-0.00015}$  & $0.02262^{+0.00016}_{-0.00014}$ & $0.02253^{+0.00010}_{-0.00012}$ & $0.02253^{+0.00013}_{-0.00011}$ & $0.02253^{+0.00011}_{-0.00013}$ &  $0.02259^{+0.00014}_{-0.00016}$
\\\hline
$\tau$ & $0.0538\pm 0.0070$ & $0.0532^{+0.0075}_{-0.0087}$  & $0.0594\pm 0.0074$ & $0.0501\pm 0.0052$ &  $0.0525^{+0.0052}_{-0.0064}$ &  $0.0579^{+0.0069}_{-0.0078}$ & $0.0637^{+0.0065}_{-0.0096}$
\\\hline
$H_0$ & $67.74^{+0.57}_{-0.66}$ & $68.41\pm 0.38$ & $69.43^{+0.72}_{-0.53}$ & $68.64^{+0.30}_{-0.38}$  & $68.45\pm 0.34$ &  $68.79^{+0.35}_{-0.40}$ & $68.99\pm 0.51$
\\\hline
$n_s$ & $0.9654^{+0.0035}_{-0.0042}$ & $0.9690\pm 0.0038$  & $0.9731\pm 0.0042$ & $0.9701^{+0.0029}_{-0.0033}$ & $0.9685\pm 0.0034$ &  $0.9705\pm 0.0034$ & $0.9713\pm 0.0037$
\\\hline
$\sigma_8$ & $0.8164\pm 0.0076$ & $0.8104\pm 0.0076$  & $0.8121^{+0.0065}_{-0.0080}$ & $0.8048\pm 0.0052$ &  $0.8073^{+0.0048}_{-0.0056}$ & $0.8120\pm 0.0074$ & $0.8160\pm 0.0068$
\\\hline
$\alpha$ & $<0.50$ &  $0.52\pm 0.17$ & $1.32\pm 0.18$ & $0.67^{+0.11}_{-0.16}$ & $0.25^{+0.09}_{-0.20}$ &  $0.73^{+0.11}_{-0.27}$ & $0.33^{+0.19}_{-0.23}$
\\\hline
$\beta$ & $0.0158^{+0.0067}_{-0.0120}$ & $0.0206^{+0.0070}_{-0.0095}$  & $0.0294^{+0.0120}_{-0.0076}$  & $0.0151^{+0.0073}_{-0.0083}$ & $0.0095^{+0.0030}_{-0.0087}$ &  $0.0206^{+0.0076}_{-0.0100}$ & $0.0197^{+0.0094}_{-0.0084}$
\\\hline
$\chi^2_{min,{\rm CDE}}-\chi^2_{min,\Lambda}$ & $-0.02$ & $-0.28$ & $-0.58$ & $-1.56$ & $-0.90$ & $-1.34$ & $-1.46$
\\\hline
$\ln\,B_{{\rm CDE},\Lambda}$ & $-8.05$ & $-9.95$ & $-7.57$ & $-8.33$ & $-7.83$ & $-7.95$ & $-8.75$
\\\hline
 \end{tabular}}
\caption{\scriptsize{Constraints obtained using the data set combinations described in \autoref{sect:CombiData} on the following parameters of the CDE model: the reduced DM and baryon energy densities, $\Omega^{(0)}_{dm}h^2$ and $\Omega^{(0)}_{b}h^2$; the reionization optical depth, $\tau$; the Hubble parameter, $H_0$ (in units of km/s/Mpc); the power of the primordial power spectrum, $n_s$; the current amplitude of mass fluctuations at $8h^{-1}\,{\rm Mpc}$, $\sigma_8$; the coupling strength $\beta$; and the power of the PR potential \eqref{eq:PeeblesRatra}. We remark that these are not the primary parameters that are varied in the Monte Carlo analyses (cf. appendix A for details). We provide the mean values and $68\%$ confidence intervals for each of them. In the last two rows we show the differences {\it w.r.t.} the $\Lambda$CDM of the minimum values of the $\chi^2$-function, and also the natural logarithm of the Bayes ratio $B_{{\rm CDE},\Lambda}$, as defined in \eqref{eq:evidence}-\eqref{eq:BayesRatio}. The (small) negative values of $\chi^2_{min.{\rm CDE}}-\chi^2_{min,\Lambda}$ tell us that CDE is able to fit slightly better the data than the $\Lambda$CDM; if we use as an alternative estimator the Bayes factor, we find negative values of $\ln(B_{{\rm CDE},\Lambda})$, indicating a preference for the $\Lambda$CDM model. See \autoref{sect:results} for a thorough discussion of the results.}}
\label{tab:tab1}
\end{center}
\end{table*}

The internal correlations between the BAO and RSD data from \cite{Gil-Marin:2016wya} and \cite{Gil-Marin:2018cgo} have been duly taken into account through the corresponding covariance matrices provided in these two references.

%%%%%%%%%%%%%%%%%%%%%%%%%%%%%%%%%%%%%%%%%%%%%%%%%%%%%%%%%%%%%%%%%

\subsubsection{Prior on $H_0$}\label{sect:H0}

In some of our data set combinations (cf. \autoref{sect:CombiData}) we include the prior on the Hubble parameter
\begin{equation}
H_{0,{\rm SH0ES}}=(74.03\pm 1.42) \,{\rm km/s/Mpc},
\end{equation}
reported by the SH0ES Team in \cite{Riess:2019cxk}. It is obtained from the cosmic distance ladder and using an improved calibration of the Cepheid period-luminosity relation, based on distances obtained from detached eclipsing binaries located in the Large Magellanic Cloud, masers in the galaxy NGC $4258$ and Milky Way parallaxes. This value of the Hubble parameter is in $4.4\sigma$ tension\footnote{The tension (in terms of the number of $\sigma$) between two quantities $A\pm\sigma_A$ and $B\pm\sigma_B$ is {\it estimated} in this work by using the formula $|A-B|/\sqrt{\sigma_A^2+\sigma_B^2}$, which strictly speaking is only valid if the two values are normally distributed and independent.} with the TTTEEE+lowE+lensing best-fit $\Lambda$CDM model of {\it Planck} 2018 \cite{Aghanim:2018eyx}, $H_0=67.36\pm 0.54$ km/s/Mpc.

It has been recently argued in \cite{Camarena:2019rmj} (and later on also in \cite{Dhawan:2020xmp,Benevento:2020fev}) that in cosmological studies it is better to use the SH0ES constraint on the absolute magnitude of the SNIa rather than the direct prior on $H_0$ when combined with low-redshift SNIa data. This is to avoid double counting issues. We do not have this problem, though, since we do not combine the Pantheon compilation with the prior from SH0ES in any of our main analyses (cf. \autoref{sect:CombiData}).

%%%%%%%%%%%%%%%%%%%%%%%%%%%%%%%%%%%%%%%%%%%%%%%%%%%%%%%%%%%%%%%%%

\subsubsection{Strong-lensing time delay distances}\label{sect:RSD}
In combination with the prior on $H_0$ from SH0ES we also use the angular diameter distances reported by the H0LICOW collaboration\footnote{http://shsuyu.github.io/H0LiCOW/site/}. They analyze six gravitationally lensed quasars of variable luminosity. After measuring the time delay between the deflected light rays and modeling the lenses they are able to measure the so-called time-delay distances $D_{\Delta t}$ (cf. \cite{Wong:2019kwg} and references therein). We use their reported six time-delay distances (one for each lensed system), and one distance to the deflector B1608+656, which according to the authors of \cite{Wong:2019kwg} is uncorrelated with the corresponding $D_{\Delta t}$. The relevant information for building the likelihood can be found in Tables 1 and 2 of \cite{Wong:2019kwg}, and their captions. Assuming the concordance model, these distances lead to a value of $H_0=(73.3^{+1.7}_{-1.8})$ km/s/Mpc, which is in $3.2\sigma$ tension with the one obtained from the TTTEEE+lowE+lensing analysis by {\it Planck} \cite{Aghanim:2018eyx}.

%%%%%%%%%%%%%%%%%%%%%%%%%%%%%%%%%%%%%%%%%%%%%%%%%%%%%%%%%%%%%%%%%
%%%%%%%%%%%%%%%%%%%%%%%%%%%%%%%%%%%%%%%%%%%%%%%%%%%%%%%%%%%%%%%%%

\subsection{Combined data sets}\label{sect:CombiData}

We proceed now to describe the data set combinations under which we have obtained the main results of this work. They are discussed in detail in \autoref{sect:results}. We put constraints using the following combinations: (i) TTTEEE+lowE CMB data from {\it Planck} 2018 \cite{Aghanim:2018eyx}, in order to see the constraining power of the CMB when used alone, and also to check whether these data lead to a higher value of $H_0$ than in the $\Lambda$CDM. For simplicity, we will refer to this data set as P18 throughout the paper; (ii) P18+BSC, with BSC denoting the background data set BAO+SNIa+CCH; (iii) We add on top of the latter the linear structure formation information contained in the RSD data, P18+BSC+RSD; (iv) We study the impact of the CMB lensing by also adding the corresponding likelihood, P18lens+BSC+RSD; (v) Finally, we analyze the impact of the prior on $H_0$ from SH0ES \cite{Riess:2019cxk} and the H0LICOW angular diameter distances \cite{Wong:2019kwg} by using the data sets P18+SH0ES+H0LICOW, P18lens+SH0ES+H0LICOW and P18+BSC+SH0ES+H0LICOW. The distance ladder and strong-lensing time delay measurements of the Hubble constant are completely independent (see e.g. the reviews \cite{Verde:2019ivm,Riess:2020sih}). When combined, they lead to 
\begin{equation}\label{eq:H0comb}
H_{0,comb}=(73.74\pm 1.10)\,{\rm km/s/Mpc}\,,
\end{equation}
in $5.2\sigma$ tension with the best-fit $\Lambda$CDM value reported by {\it Planck} 2018 \cite{Aghanim:2018eyx}. Hence, it is interesting to check what is the response of the CDE model under these concrete data sets, and to compare the results with those obtained using only the CMB likelihood.

%%%%%%%%%%%%%%%%%%%%%%%%%%%%%%%%%%%%%%%%%%%%%%%%%%%%%%%%%%%%%%%%%
%%%%%%%%%%%%%%%%%%%%%%%%%%%%%%%%%%%%%%%%%%%%%%%%%%%%%%%%%%%%%%%%%

%%%%%%%%%%%%%%%%%%%%%%%%%%%%%%%%%%%%%%%%%%%%%%%%%%%%%%%%%%%%%%%%%
%%%%%%%%%%%%%% FIGURE 3 %%%%%%%%%%%%%%%%%%%%%%%%%%%
%%%%%%%%%%%%%%%%%%%%%%%
%%%%
\begin{figure*}
\begin{center}
\includegraphics[width=6in, height=3.1in]{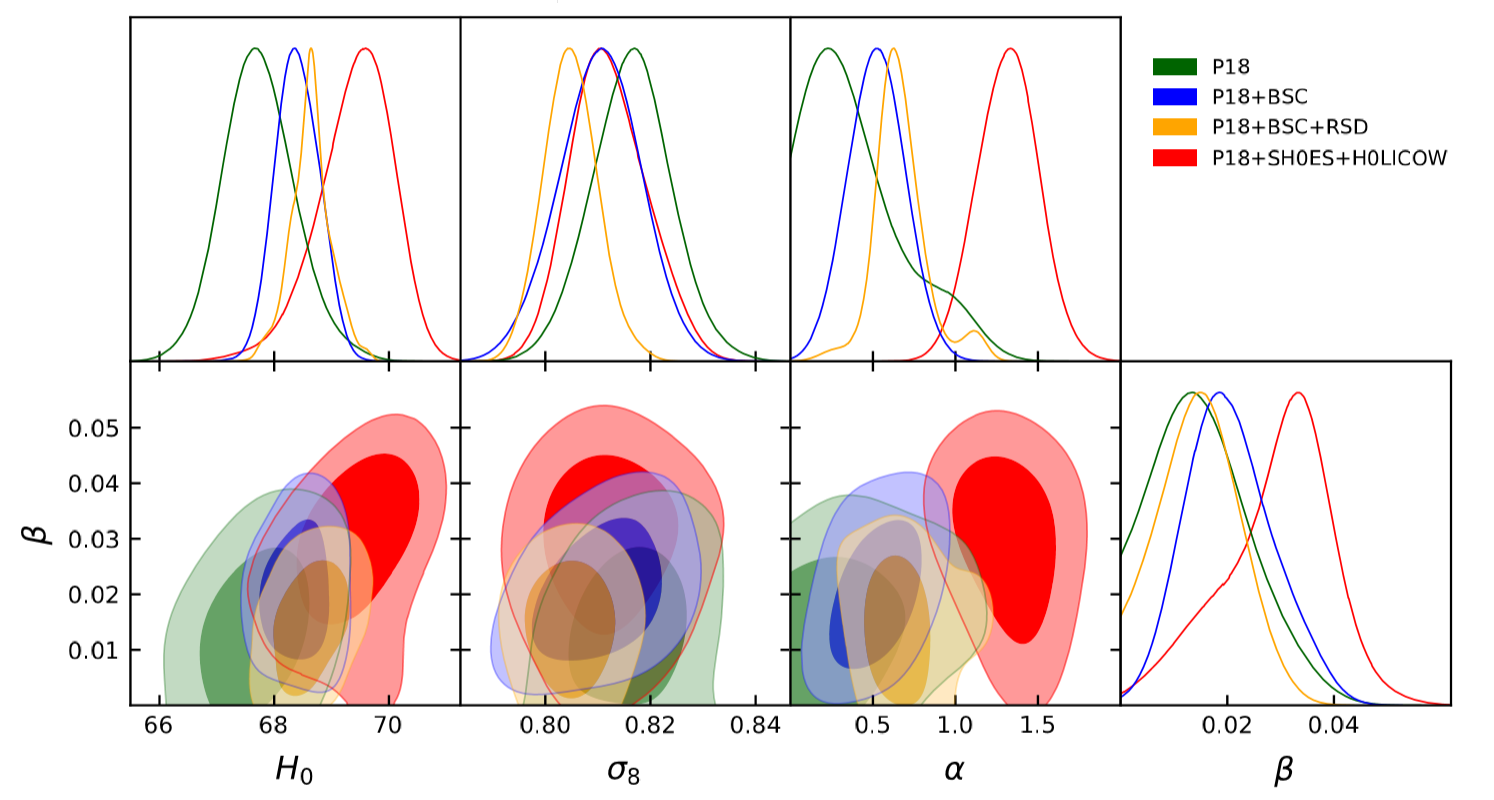}
\caption{\scriptsize {$1$ and $2\sigma$ confidence contours obtained using some of the combined data sets described in \autoref{sect:CombiData} in the $(H_0,\beta)$, $(\sigma_8,\beta)$, and $(\alpha,\beta)$ planes, together with the marginalized one-dimensional posterior distributions for these parameters. See the discussion of these results in \autoref{sect:results}.}}
\label{fig:betaCLs}                                                      
\end{center}
\end{figure*}
%%%%%%
%%%%%%%%%%%%%%%%%%%%%%%%%%%%%%%%%%%%%%%%%%%%%%%%%%%%%%%%%%%%%%%%%
%%%%%%%%%%%%%%%%%%%%%%%%%%%%%%%%%%%%%%%%%%%%%%%%%%%%%%%%%%%%%%%%%

\section{Results}\label{sect:results}

\begin{figure*}
\begin{center}
\includegraphics[width=4.5in, height=4.5in]{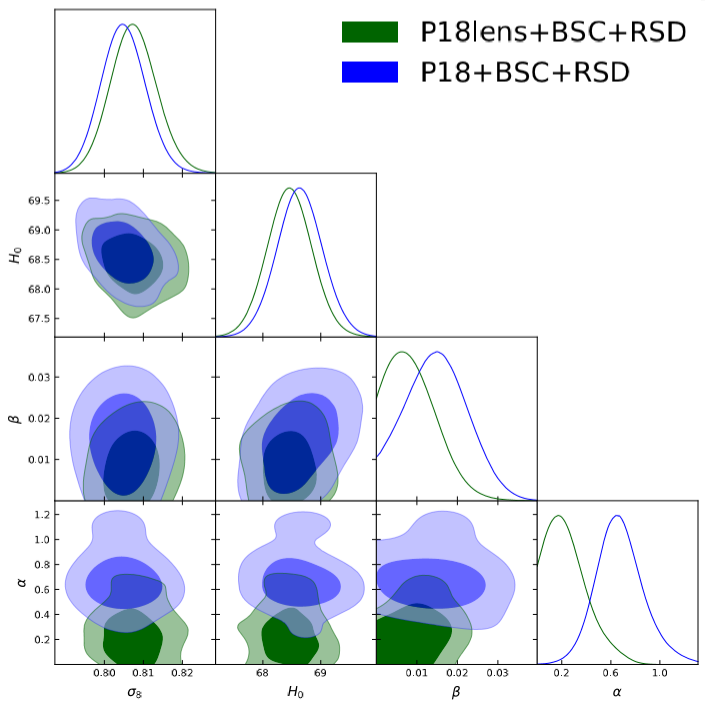}
\caption{\scriptsize {$1$ and $2\sigma$ confidence contours obtained with the P18+BSC+RSD and P18lens+BSC+RSD data sets in the most relevant two-dimensional planes of the CDE model parameter space. They allow us to see what is the impact of the CMB lensing on our results. We also show the corresponding marginalized one-dimensional posterior distributions for all the parameters. See the related comments in \autoref{sect:results}.}}
\label{fig:nolensVSlens_reduced}                                                      
\end{center}
\end{figure*}

Our main results are presented in Tables \ref{tab:tab1}-\ref{tab:tab2} and Figs. \ref{fig:betaCLs}-\ref{fig:nolensVSlens_reduced}.  When we only employ the CMB temperature and polarization data from {\it Planck} 2018 \cite{Aghanim:2018eyx} (i.e. the P18 data set) to constrain the CDE model, the fitting values obtained for $\alpha$ and $\beta$ are compatible at $1\sigma$ c.l. with 0, i.e. with a cosmological constant and no interaction in the dark sector (cf. the first column in \autoref{tab:tab1}). The value of $H_0$ remains low, roughly $4.1 \sigma$ below the cosmic distance ladder measurement of \cite{Riess:2019cxk}. Similarly, when we combine \textit{Planck} with BSC background data or with BSC+RSD, we get a value of $H_0$ which is $3.8\sigma$ and $3.7\sigma$ away from the SH0ES value, respectively.

As we have explained in \autoref{sect:CDEmodel}, there is a degeneracy between the strength of the fifth force, i.e. the parameter $\beta$, and the Hubble parameter. CDE is in principle able to lower the value of the sound horizon at the decoupling time, $r_s$, and the amplitude of the first peak of the $\mathcal{D}^{TT}_{l}$'s. The CMB data fix with high precision the angle $\theta_*=r_s/D^{(c)}_A(z_{dec})$, with $D^{(c)}_A(z_{dec})$ the comoving angular diameter distance to the CMB last scattering surface. This means that in order to keep this ratio constant, $H_0$ will tend to grow for increasing values of the coupling strength, so that $D^{(c)}_A(z_{dec})$ decreases and compensates in this way the lowering of $r_s$, while keeping the height of the first peak compatible with data. This positive correlation between $H_0$ and $\beta$ can be appreciated in the left-most contour plot of \autoref{fig:betaCLs}. The latter shows 1 and 2$\sigma$ posterior probabilities for a selection of cosmological parameters. As discussed, we confirm from the first plot a mild degeneracy between $H_0$ and $\beta$. The strength of the fifth force does not seem to be very degenerate with $\sigma_8$ nor with the potential parameter $\alpha$.

%%%%%%%%%%%%%%%%%%%%%%%%%%%%%%%%%%%%%%%%%%%%%%%%%%%%%%%%%%%%%%%%%
%%%%%%%%%%%%%% FIGURE 4 %%%%%%%%%%%%%%%%%%%%%%%%%%%
%%%%%%%%%%%%%%%%%%%%%%%
%%%%
%%%%%%
%%%%%%%%%%%%%%%%%%%%%%%%%%%%%%%%%%%%%%%%%%%%%%%%%%%%%%%%%%%%%%%%%
%%%%%%%%%%%%%%%%%%%%%%%%%%%%%%%%%%%%%%%%%%%%%%%%%%%%%%%%%%%%%%%%%

Impact of adding background data on top of P18 can be grasped by looking at the one-dimensional posterior distributions of \autoref{fig:betaCLs} (in blue), and also at the numbers of the second column of \autoref{tab:tab1}. Using the P18+BSC combined data set we find that $\beta$ and $\alpha$ are now $\sim 2.5$ and $\sim 3.1\sigma$ away from 0, respectively. The values of $H_0$ and $\sigma_8$, are however compatible at $1\sigma$ with the ones obtained using only the P18 data set. They are also fully compatible with those obtained with the $\Lambda$CDM under the same data set, which read: $H_0=(68.29\pm 0.37)$ km/s/Mpc, $\sigma_8=0.812^{+0.006}_{-0.008}$. The peaks in $\beta$ and $\alpha$ may indicate a mild preference of low-redshift data, when combined with the CMB, for a non-null interaction in the dark sector and a running quintessence potential. As noted already in \cite{Ade:2015rim}, we remark that this preference does not seem to correspond to a large improvement in the minimum value of $\chi^2$ with respect to the $\Lambda$CDM: under the P18+BSC data set, $\chi^2_{min,{\rm CDE}}-\chi^2_{min,\Lambda}$ is negative, but very close to 0, which means that the CDE model only is able to improve the description of the data in a very marginal way. 

The addition of the RSD data to the P18+BSC combined data set doesn't change much the result: there is a very small shift in the peak of the one-dimensional posterior distribution for $\alpha$ to larger values and the one for $\beta$ to lower ones (see the yellow curves in \autoref{fig:betaCLs}). These two facts reduce a little bit the value of $\sigma_8$. The aforesaid peaks are now $\sim 5$ and $\sim 2\sigma$ away from 0, respectively, with a reduction of $\chi^2_{min}$ {\it w.r.t.} the $\Lambda$CDM of 1.56 units (cf. \autoref{tab:tab1}, fourth column), i.e. pointing to a very small preference for CDE. The value of $H_0$ is almost unchanged.

If we include also the CMB lensing information, i.e. if we consider the P18lens+BSC+RSD combined data set, posterior probabilities squeeze, as expected, towards the $\Lambda$CDM values. This can be seen in \autoref{fig:nolensVSlens_reduced}, and also in the fifth column of \autoref{tab:tab1}. Given the caveats explained in \autoref{sect:CMB}, we find important to highlight the specific impact of CMB lensing data with respect to the P18+BSC+RSD data set. 

In order to further evaluate the level at which the degeneracy observed in the ($H_0$,$\beta$)-plane can alleviate the tension in the Hubble parameter between \textit{Planck} and $\{$SH0ES, H0LICOW$\}$ data, we perform a Monte Carlo analysis combining those data within the CDE model: results are shown in the third column in \autoref{tab:tab1} and correspond to red contours in \autoref{fig:betaCLs}. In this case, the best fit corresponds to a value of $\beta =0.0294^{+0.0120}_{-0.0076}$, i.e. at $3\sigma$ from zero coupling, a value of $\alpha =1.32\pm 0.18$, with $\alpha > 0$ at $\sim 7\sigma$ c.l., and $H_0 =(69.43^{+0.72}_{-0.53})$ km/s/Mpc. The raise of $H_0$ is possible thanks to the increase of $\beta$, which in turn needs also larger values of $\alpha$. The tension with the SH0ES+H0LICOW measurement \eqref{eq:H0comb} is slightly reduced from $4.8\sigma$ (when only P18 is used to constrain the model, cf. the first column of \autoref{tab:tab1}) to $3.5 \sigma$ (when also the SH0ES+H0LICOW data are considered). This shifts the $H_0$ value $1.9\sigma$ higher than the best fit using the P18 data set alone, within CDE. Combining also with background data, such as BSC, can partially break degeneracies and leads to  $\alpha =0.73^{+0.11}_{-0.27}$, with $\alpha > 0$ at $3.8\sigma$ and $H_0=(68.79^{+0.35}_{-0.40})$ km/s/Mpc at $4.3\sigma$ from the SH0ES+H0LICOW value \eqref{eq:H0comb}, reducing the chance of CDE to alleviate the tension, as shown in the penultimate column of the table. Finally, the impact of adding CMB lensing is shown in the last column, where now $\beta=0.0197^{+0.0094}_{-0.0084}$ and $\alpha =0.33^{+0.19}_{-0.23}$, with $\beta>0$ and $\alpha > 0$ at $2.2\sigma$ and $1.6\sigma$, respectively, i.e. shifting back towards $\Lambda$CDM. In this case $H_0=(68.99\pm 0.51)$ km/s/Mpc, $3.9\sigma$ away from the SH0ES+H0LICOW value \eqref{eq:H0comb} and even more had we included also BSC. 

Finally, we can further quantify the relative ability of the CDE model to describe the various data sets {\it w.r.t.} the $\Lambda$CDM cosmology using the Bayes ratio, in alternative to the more approximate $\chi^2$ estimate we mentioned so far.  Given a data set $\mathcal{D}$, the probability of a certain model $M_i$ to be the best one among a given set of models $\{M\}$ reads, 
\begin{equation}\label{eq:BayesTheorem}
P(M_i|\mathcal{D})=\frac{P(M_i)\mathcal{E}(\mathcal{D}|M_i)}{P(\mathcal{D})}\,,
\end{equation}
where $P(M_i)$ is the prior probability of the model $M_i$ and  $P(\mathcal{D})$ the probability of having the data set $\mathcal{D}$. Obviously, the normalization condition $\sum_{j\in\{M\}}P(M_j)=1$ must be fulfilled. The quantity $\mathcal{E}(\mathcal{D}|M_i)$ is the so-called marginal likelihood or evidence. If the model $M_i$ has $n$ parameters $p^{M_i}_1, p^{M_i}_2,...,p^{M_i}_n$, the evidence takes the following form, 
\begin{equation}\label{eq:evidence}
\mathcal{E}(\mathcal{D}|M_i)=\int \mathcal{L}(\mathcal{D}|\vec{p}^{M_i},M_i)\pi(\vec{p}^{M_i}) d^np^{M_i}\,,
\end{equation}
with $\mathcal{L}(\mathcal{D}|\vec{p}^{M_i},M_i)$ being the likelihood and $\pi(\vec{p}^{M_i})$ the prior of the parameters entering the model $M_i$. The evidence is larger for those models that have more overlapping volume between the likelihood and the prior distributions, but penalizes the use of additional parameters having a non-null impact on the likelihood. Hence, the evidence constitutes a good way of quantifying the performance of the model by implementing in practice the Occam razor principle. If we compare the CDE and $\Lambda$CDM models by assuming equal prior probability for both of them, i.e. $P({\rm CDE})=P(\Lambda {\rm CDM})$, then we find that the ratio of their associated probabilities is directly given by the ratio of their corresponding evidences, i.e.
\begin{equation}\label{eq:BayesRatio}
\frac{P({\rm CDE}|\mathcal{D})}{P(\Lambda {\rm CDM}|\mathcal{D})}=\frac{\mathcal{E}(\mathcal{D}|{\rm CDE})}{\mathcal{E}(\mathcal{D}|\Lambda {\rm CDM})} \equiv B_{{\rm CDE},\Lambda}\,.
\end{equation}
This is known as Bayes ratio and is the quantity we are interested in. For more details we refer the reader to \cite{KassRaftery1995,Burnham2002,AmendolaTsujikawaBook}. Notice that the computation of \eqref{eq:BayesRatio} is not an easy task in general, since  we usually work with models with a high number of (mostly nuisance) parameters, so the integrals under consideration becomes quite involved. We have computed the evidences numerically using the Markov chains obtained from the Monte Carlo analyses and with the aid of the numerical code \texttt{MCEvidence} \cite{Heavens:2017afc}, which is publicly available (cf. \autoref{sect:methodology}). We report the values obtained for the natural logarithm of the Bayes ratio \eqref{eq:BayesRatio} in the last row of \autoref{tab:tab1}. For all the data sets under study we find values of  $\ln(B_{{\rm CDE},\Lambda})<-5$, which point to a preference of the $\Lambda$CDM over the CDE model according to Jeffreys' scale \cite{KassRaftery1995,Burnham2002,AmendolaTsujikawaBook}. Although the CDE model we are studying here is able to reduce slightly the value of $\chi^2_{min}$ {\it w.r.t.} the $\Lambda$CDM, it has two additional parameters, namely $\alpha$ and $\beta$. Moreover, the initial value of the scalar field, $\phi_{ini}$, is also left free in the Monte Carlo analysis, cf. Appendix A for details\footnote{In the computation of the evidence \eqref{eq:evidence} for the CDE model we have employed the following flat priors for the extra parameters: $0<\beta<0.1$, $0<\alpha<2$, and $0<\kappa\phi_{ini}<50$. Slightly broader or tighter priors can be considered, but $\ln(B_{{\rm CDE},\Lambda})$ only changes logarithmically, so our conclusions are not very sensitive to them.}. It turns out that the decrease in $\chi^2_{min}$ is insufficient to compensate the penalization introduced by the use of these extra parameters. If instead of using the evidences \eqref{eq:evidence} and the Bayes ratio \eqref{eq:BayesRatio} to perform the model comparison we make use of e.g. the Akaike \cite{Akaike}, Bayesian \cite{Schwarz1978} or Deviance \cite{DIC} information criteria, we reach similar conclusions\footnote{For instance, Akaike criterion \cite{Akaike} is given by ${\rm AIC}= \chi^2_{min}+2n$, where $n$ is the number of parameters in the model (the degree of correlation between them is not taken into account). Considering that CDE with PR potential has an effective number of parameters between 2 and 3 we find $2.5<{\rm AIC}_{\rm CDE}-{\rm AIC}_\Lambda<6$ for the scenarios explored in \autoref{tab:tab1}, which leads to a positive preference for $\Lambda$CDM, again using Jeffreys' scale \cite{KassRaftery1995,Burnham2002,AmendolaTsujikawaBook}.}. We want to note, though, that all these information criteria are approximations of the exact Bayesian approach. Although they allow to skip the demanding computation of the  evidence  \eqref{eq:evidence}, they are only reliable when the posterior distribution is close to a multivariate Gaussian (which is not the case under study), and the Akaike and Bayesian criteria do not take into account the impact of priors nor the existing correlations between the parameters. 

Similar results and conclusions are reached using an exponential potential for the scalar field, instead of \eqref{eq:PeeblesRatra}. They are presented and discussed in appendix C.

Finally, our results are compatible with the ones in \cite{2020PhRvD.101f3502D}: the inclusion of background and CMB lensing shifts constraints towards $\Lambda$CDM; the model is however different, as ours starts from modifying the Lagrangian, which is not available in \cite{2020PhRvD.101f3502D}; furthermore, the source function is also different and while the DE EoS parameter $w$ has to be fixed to a very specific value in \cite{2020PhRvD.101f3502D} in order to match stability conditions specific to that scenario, in our case it varies; the extra parameters leads then to a more negative Bayes ratio, preferring $\Lambda$CDM, more than claimed in \cite{2020PhRvD.101f3502D}.

%%%%%%%%%%%%%%%%%%%%%%%%%%%%%%%%%%%%%%%%%%%%%%%%%%%%      
%%%%%%%%%%%%%%%%%%%%%%%%%%%%%%%%%%%%%%%%%%%%%%%%%%%%
%%%%%%%%%%%%%%%%%%%%%%%%%%%%%%%%%%%%%%%%%%%%%%%%%%%%

\section{Conclusions}\label{sect:conclusions}

Cosmological observations help to test the dark sector, and in particular   interactions between dark matter particles, mediated by a dark energy scalar field, as in CDE cosmologies. Up to a conformal transformation, this is another way of testing gravity at large scales. In this paper we carried out this task in one of the simplest and most studied models, namely, a dark energy-dark matter conformal coupling with a Peebles-Ratra potential. CDE might probe helpful to explain the well-known tension between local and cosmological values of $H_0$. Any detection of a varying dark energy potential or interaction would clearly constitute a major result and it is therefore important to monitor the constraints that newer data impose. This is particularly true in view of earlier results that detected a non-zero value of the coupling $\beta$ \cite{Pettorino:2013oxa, Ade:2015rim}.

We confirm the existence of a peak in the marginalized posterior distribution for $\beta$ and $\alpha$, more or less evident depending on the data set combination. While for P18 + SH0ES + H0LICOW $\beta > 0$ at $3\sigma$ and $\alpha > 0$ at nearly $7 \sigma$, inclusion of background data reduces the evidence to $\beta > 0$ at $2.3\sigma$ and $\alpha > 0$ at nearly $3.8 \sigma$. Inclusion of CMB lensing shifts both values to be compatible with $\Lambda$CDM within 2$\sigma$. We find it important to stress that specifically CMB lensing prefers $\Lambda$CDM, and recalled in \autoref{sect:CMB} the caveats that would deserve further investigation in order to make this result robust also for models that depart from $\Lambda$CDM as much as CDE. In all cases, we find that, overall, the peak does not correspond to a better Bayes ratio and $\Lambda$CDM remains the favored model when employing Bayesian model comparison, given the extra parameters introduced within the model. With regard to $H_0$, we find that under the P18+SH0ES+H0LICOW combined data set the simple coupled model with constant coupling investigated in this work leads to a value in $3.5\sigma$ tension with \eqref{eq:H0comb}, or in 4.3$\sigma$ tension when including further background data. The values of $\sigma_8$ are also similar to those found in the $\Lambda$CDM (i.e. $\sigma_8\sim 0.80-0.82$), even when RSD data are considered together with CMB and background data. In this case we find $\beta=0.010^{+0.003}_{-0.009}$ and $\beta=0.015^{+0.007}_{-0.008}$, with and without CMB lensing, respectively. For the values of the coupling strength preferred by the data we find the typical increase of the mass of the DM particles to be $m(\phi_{ini})/m^{(0)}-1\lesssim \mathcal{O}(1)\%$.

The question that naturally arises is then, which modification of CDE can help alleviating the tensions? One can immediately suppose that a varying $\beta$ can go some way towards this. Or, it could be that a model with both energy- and momentum-couplings (see e.g. \cite{Amendola:2020ldb}), which can introduce a weaker gravity, helps with the tensions. These issues will be investigated in future publications.

%%%%%%%%%%%%%%%%%%%%%%%%%%%%%%%%%%%%%%%%%%%%%%%%%%%%
%%%%%%%%%%%%%%%%%%%%%%%%%%%%%%%%%%%%%%%%%%%%%%%%%%%%
%%%%%%%%%%%%%%%%%%%%%%%%%%%%%%%%%%%%%%%%%%%%%%%%%%%%

\vspace{1cm}

{\bf Acknowledgements}
\newline
\newline
\noindent
This paper was completed during the COVID-19 pandemic. We thank our institutions for allowing us to work remotely.
AGV is funded by the Deutsche Forschungsgemeinschaft (DFG) - Project number 415335479.

%%%%%%%%%%%%%%%%%%%%%%%%%%%%%%%%%%%%%%%%%%%%%%%%%%%%%%%%%%%%%%%%%
%%%%%%%%%%%%%%%%%%%%%%%%%%%%%%%%%%%%%%%%%%%%%%%%%%%%%%%%%%%%%%%%%
%%%%%%%%%%%%%%%%%%%%%%%%%%%%%%%%%%%%%%%%%%%%%%%%%%%%%%%%%%%%%%%%%

\appendix 

\section*{Appendix A: Avoiding the shooting}

IDEA takes as input parameters the current energy densities of the various species and applies a shooting method (see e.g. \cite{StoerBulirsch1980}) to find the initial energy densities that lead to the present-day values specified in the input. This is of course a very useful and convenient way of implementing the model, since very often we are interested in computing theoretical quantities by fixing the current energy densities to concrete values, most of the times very close to the best-fit $\Lambda$CDM ones. This trial and error method is unavoidable if one wants to do so. Nevertheless, this is not the most efficient way to proceed at the level of the Monte Carlo analysis. The avoidance of the shooting recursive process by directly using as input parameters the initial conditions of the energy densities instead of their current values allows us to save precious computational time. In our implementation of the CDE model in \texttt{CLASS} \cite{Blas:2011rf} we have skipped the shooting method proceeding in this way. The current energy densities and other quantities of interest, e.g. $H_0$ or $\sigma_8$, are obtained as derived parameters after solving the complete set of Einstein-Boltzmann equations up to $a=1$. We also use as input parameter in the Monte Carlo the initial value of the scalar field, $\phi_{ini}=\phi(a_{ini})>0$, with $a_{ini}=10^{-14}$. On the contrary, $\phi^\prime_{ini}=\phi^\prime(a_{ini})$ can be expressed in terms of other input parameters. Let us show how. By solving \eqref{eq:KGeq} in the radiation-dominated epoch (RDE) we find,
%
%\begin{equation}\label{eq:phiPrimeConst}
%\phi^\prime(\tau)=150 \beta\, \frac{\Omega_{dm}(a_{ini})}{\kappa a_{ini}}\,\sqrt{\omega^{*}_r}\times({\rm km/s/Mpc})+\frac{c_0}{\tau^2}\,,
%\end{equation}
\begin{equation}\label{eq:phiPrimeConst}
\phi^\prime(\tau)=150 \beta\, \frac{\Omega_{dm}(a_{ini})}{\kappa a_{ini}}\,\varsigma\sqrt{\omega^{*}_r}+\frac{c_0}{\tau^2}\,,
\end{equation}
where $c_0$ is a dimensionless integration constant, $\varsigma\equiv 1\,{\rm km/s/Mpc}=2.1332\times 10^{-44}$ GeV (in natural units), and $\omega^{*}_r=\omega_\gamma(1+0.2271 N_{eff})$ is the reduced density parameter of radiation during the RDE. We consider three massive neutrinos with equal mass and $\sum_\nu m_\nu=0.06$ eV, so $N_{\rm eff}=3.046$. The parameter $\omega_\gamma$ is determined by the temperature of the CMB photons at present, which we set to the value reported in \cite{Fixsen:2009ug}, $T^{(0)}_{\gamma}=2.7255\,K$. Notice that the ratio $\Omega_{dm}(a)/a$ appearing in \eqref{eq:phiPrimeConst} is kept constant during the RDE. To understand this let us consider equation \eqref{eq:consDM}. It can be rewritten as
\begin{equation}
\rho_{dm}^\prime+3\mathcal{H}\rho_{dm}\left(1-\beta\sqrt{\frac{2}{3}\Omega_{\phi,{\rm kin}}(a)}\right)=0\,,
\end{equation}
%
%
%%%%%%%%%%%%%%%%%%%%%%%%%%%%%%%%%%%%%%%%%%%%%%%%%%%%%%%%%%%%%%%%%%%%%%%%%%%%%%
%
\begin{table*}[t]
\begin{center}
\resizebox{11cm}{!}{
%\begin{scriptsize}
\begin{tabular}{| c | c |c | c | c|}
\multicolumn{1}{c}{Parameter}  &  \multicolumn{1}{c}{$\Lambda$CDM} & \multicolumn{1}{c}{{\small Peebles-Ratra}} & \multicolumn{1}{c}{CDE with $\alpha=0$} &  \multicolumn{1}{c}{CDE}
\\\hline
$\Omega^{(0)}_{dm}h^2$ & $0.1188\pm 0.0008$ & $0.1180^{+0.0010}_{-0.0009}$  & $0.1187^{+0.0006}_{-0.0008}$ & $0.1187\pm 0.0008$
\\\hline
$\Omega^{(0)}_{b}h^2$  & $0.02252\pm 0.00012$ & $0.02257\pm 0.00014$ & $0.02253\pm 0.00011$ & $0.02253^{+0.00010}_{-0.00012}$
\\\hline
$\tau$  & $0.0508^{+0.0048}_{-0.0072}$ & $0.0532^{+0.0063}_{-0.0079}$ & $0.0496\pm 0.0047$ &  $0.0501\pm 0.0052$
\\\hline
$H_0$ [km/s/Mpc]  & $68.50\pm 0.34$ & $67.68^{+0.61}_{-0.52}$ & $68.55^{+0.38}_{-0.31}$ & $68.64^{+0.30}_{-0.38}$
\\\hline
$n_s$  & $0.9696\pm 0.0034$ & $0.9719\pm 0.0038$  & $0.9700^{+0.0032}_{-0.0037}$ & $0.9701^{+0.0029}_{-0.0033}$
\\\hline
$\sigma_8$  & $0.8033\pm 0.0057$ & $0.7880^{+0.0110}_{-0.0097}$ & $0.8022\pm 0.0054$ & $0.8048\pm 0.0052$
\\\hline
$\alpha$  & - & $0.096^{+0.038}_{-0.071}$ & - & $0.67^{+0.11}_{-0.16}$
\\\hline
$\beta$  & - & - & $0.0040^{+0.0012}_{-0.0032}$ &   $0.0151^{+0.0073}_{-0.0083}$
\\\hline
$\chi^2_{min,i}-\chi^2_{min,\Lambda}$  & - & $-1.74$ & $-1.02$ & $-1.56$ %chi^2_{min,L} = 2821.70
\\\hline
$\ln\,B_{i,\Lambda}$ & - & $-1.67$ & $-5.14$ & $-8.33$
\\\hline
 \end{tabular}}
\caption{\scriptsize{Constraints for the $\Lambda$CDM, PR, CDE with flat potential and general CDE models obtained using the P18+BSC+RSD data set. See the comments in Appendix B.}}
\label{tab:tab2}
\end{center}
\end{table*}
%%%%%%%%%%%%%%%%%%%%%%%%%%%%%%%%%%%%%%%%%%%%%%%%%%%%%%%%%%%%%%%%%%%%%%%%%%%%%%
with $\Omega_{\phi,{\rm kin}}$ being the fraction of scalar field kinetic energy in the Universe. During the RDE $\Omega_{\phi,{\rm kin}}\sim 0$. In addition, $\beta\ll 1$, so we find that $\rho_{dm}\sim a^{-3}$ and, hence, $\Omega_{dm}(a)/a=const.=\Omega_{dm}(a_{ini})/a_{ini}$. The first term in the {\it r.h.s.} of \eqref{eq:phiPrimeConst} is, therefore, constant. The solution \eqref{eq:phiPrimeConst} does not depend on the form of the scalar field potential, since the impact of the latter is completely negligible during the RDE, and consists of a constant term plus a fast decaying mode, which we will call $\phi^\prime_{\rm cons}$ and $\phi^\prime_{\rm dec}$, respectively. In order to fulfill the BBN constraint on the total energy density at $a_{\rm BBN}\sim 10^{-9}$ one needs to demand $\rho_\phi(a_{\rm BBN})\lesssim 0.1\cdot\rho_r(a_{\rm BBN})$ \cite{Uzan:2010pm}. This leads to the following condition: $|c_0|<10^{53}$. Now, using the value of $c_0$ that saturates the upper bound we can evaluate the ratio $\phi^\prime_{\rm dec}(a)/\phi^\prime_{\rm cons}(a)$ at any moment of the RDE (knowing that $a(\tau)=100\,\tau\,\varsigma\, \sqrt{\omega_r^{*}}$). In particular, we can compute it at a moment near the end of the RDE, e.g. at $\tilde{a}=10^{-4}$, and see whether the decaying mode can still play an important role at that time. If we do so we obtain $\phi^\prime_{\rm dec}(\tilde{a})/\phi^\prime_{\rm cons}(\tilde{a})\approx 10^{-5}/\beta$. The values of the coupling strength explored in our Monte Carlo analyses are in the range $10^{-3}\lesssim\beta\lesssim 10^{-1}$, so we find
\begin{equation}
10^{-4}\lesssim\frac{\phi^\prime_{\rm dec}(\tilde{a})}{\phi^\prime_{\rm cons}(\tilde{a})}\lesssim 10^{-2}\,.
\end{equation}
This tells us that the decaying mode will play no role in our analysis (even when $c_0$ takes the largest value allowed by the BBN condition), since the observables that we use to constrain the CDE model in this work are insensitive to $\phi^\prime$ at even lower values of the scale factor, i.e. at $a<\tilde{a}$. This is very positive because, in practice, this allows us to set the initial condition of $\phi^\prime(a_{ini})=\phi^\prime_{\rm cons}(a_{ini})$ and reduce in this way the number of parameters that are varied in each step of the Monte Carlo. This also helps to improve the efficiency of our code.

%%%%%%%%%%%%%%%%%%%%%%%%%%%%%%%%%%%%%%%%%%%%%%%%%%%%%%%%%%%%%%%%%
%%%%%%%%%%%%%%%%%%%%%%%%%%%%%%%%%%%%%%%%%%%%%%%%%%%%%%%%%%%%%%%%%
%%%%%%%%%%%%%%%%%%%%%%%%%%%%%%%%%%%%%%%%%%%%%%%%%%%%%%%%%%%%%%%%%

\section*{Appendix B: Results for the nested models}

Here we present and discuss the fitting results for the two nested models of the CDE scenario that are obtained by turning off the interaction, and also by using a constant potential while keeping active the interaction in the dark sector. These two models are obtained from the general CDE scenario described in \autoref{sect:CDEmodel} by setting $\beta=0$ and $\alpha=0$, respectively. The former corresponds to the PR model \cite{Peebles:1987ek,Ratra:1987rm}. In \autoref{tab:tab2} we show the constraints obtained for these models in the light of the P18+BSC+RSD data set, and also compare their statistical performance with the $\Lambda$CDM and the full CDE model. In practice, they both have one additional parameter {\it w.r.t.} the $\Lambda$CDM. The PR model has a very effective attractor solution for $\phi$ and $\phi^\prime$ during the radiation-dominated epoch, which can be used to fix the initial conditions of the scalar field and its derivative, so only $\alpha$ enters as an additional parameter (see e.g. \cite{Sola:2016hnq}). On the other hand, the CDE model with flat potential only has $\beta$ as extra parameter, since the equations are invariant under translations of the scalar field and hence $\phi_{ini}$ can be fixed to an arbitrary value, e.g. 0. Moreover, $\phi^\prime_{ini}$ can be set as explained in the appendix A. \autoref{tab:tab2} shows that in the context of the PR model it is possible to obtain much lower values of $\sigma_8$, loosening in this way the $\sigma_8$ tension. $H_0$, though, is below the one obtained with the $\Lambda$CDM and the other two nested models. These results are fully aligned with those from \cite{Sola:2018sjf}, but now we obtain lesser levels of evidence for the PR model, basically due to the use of the CMB high multipole polarization data, which were not employed in that reference. The reduction in the value of $\chi^2_{min}$ {\it w.r.t.} the $\Lambda$CDM is $\sim 2$ units, but $\ln(B_{{\rm PR},\Lambda})<-3$, so there is still more evidence for the concordance model when compared with the PR. One thing that we should explain is why the value of $\chi^2_{min}$ obtained with the PR model is lower than in the general CDE model. We would expect this not to happen, since the latter is an extension of the former, with two extra free parameters. The reason is the following. In our Monte Carlo analysis for the CDE model we cannot explore the region of parameter space with a pure PR behavior. In order this to happen we should produce values of $\beta$ in our chains much lower than the ones we actually produce (which are in all cases greater than $\sim 10^{-3}$ due to the flat prior on $\beta>0$ and its typical variance). These values of $\beta$ always give rise to non-completely negligible effects in the MDE, and hence there is always a departure from the pure PR model. Thus, it is not strange that we find points in parameter space of the PR model which lead to lower values of $\chi^2$ than those found in our analysis of the CDE.

The values of the parameters obtained for the CDE model with $\alpha=0$ remain very close to the $\Lambda$CDM ones (cf. the third column of \autoref{tab:tab2}). The model sticks to the $\Lambda$CDM because in this case there is no varying potential able to compensate the effects generated by the non-null coupling, so $\beta$ is forced to remain small. In terms of Occam's razor and the corresponding Bayes ratio there is a preference for the $\Lambda$CDM. The central value of $\beta$ is almost four times smaller than in the general CDE model. Something similar happens in the PR model for $\alpha$, which is now $\sim 7$ times smaller than in the general CDE scenario. Due to the fact that $\alpha$ and $\beta$ can compensate effects from each other, in the general CDE model these two parameters can be quite larger, as it is seen in the last column of \autoref{tab:tab2}.

%%%%%%%%%%%%%%%%%%%%%%%%%%%%%%%%%%%%%%%%%%%%%%%%%%%%%%%%%%%%%%%%%
%%%%%%%%%%%%%%%%%%%%%%%%%%%%%%%%%%%%%%%%%%%%%%%%%%%%%%%%%%%%%%%%%
%%%%%%%%%%%%%%%%%%%%%%%%%%%%%%%%%%%%%%%%%%%%%%%%%%%%%%%%%%%%%%%%%

\section*{Appendix C: Constraints on CDE with exponential potential}

In this brief appendix we complement the results provided in the main body of the paper, which have been obtained using the power-law potential \eqref{eq:PeeblesRatra}. In \autoref{tab:tab3} we provide constraints on CDE with the exponential potential
\begin{equation}\label{eq:EXP}
V(\phi)=V_0e^{-\lambda\kappa\phi} \,.
\end{equation}
The constant $\lambda>0$ controls its steepness. As mentioned in \autoref{sect:CDEmodel}, the quintessence potential only rules the scalar field dynamics in the late-time universe, after the $\phi$MDE epoch, when the effects coming from the interaction in the dark sector are already subdominant. Therefore, we should not expect a change in the form of the potential to affect severely the constraints on the coupling strength $\beta$, and this is actually what we find. By comparing the results provided in Tables I and III obtained under the same data sets we notice that both, the central values and uncertainties for $\beta$, are almost identical. They are also similar to the values reported in Table II of \cite{vandeBruck:2017idm}, which were obtained using the CMB likelihoods from {\it Planck} 2015, older SNIa, BAO and CCH data, and also older distance ladder priors on the Hubble parameter. Our constraints are a little bit tighter due to the updated (richer) data sets employed here. Also the values of $\lambda$ are quite similar. We note, though, that the central values of $H_0$ are mildly ($\sim 1\sigma$) lower than those obtained with the Peebles-Ratra potential. The values of $\ln\,B_{{\rm CDE},\Lambda}$ are higher (lower in absolute value) since in this model the goodness of fit is kept at the same level as in the CDE model with PR potential, and $\phi_{ini}$ plays no role and can be fixed, reducing thereby the complexity of the model. But they are still below -5. The results obtained with \eqref{eq:EXP} are hence fully consistent with those derived with \eqref{eq:PeeblesRatra}.

%
%%%%%%%%%%%%%%%%%%%%%%%%%%%%%%%%%%%%%%%%%%%%%%%%%%%%%%%%%%%%%%%%%%%%%%%%%%%%%%
%
\begin{table}
\begin{center}
\resizebox{8.2cm}{!}{
\begin{tabular}{| c | c |c | c |}
\multicolumn{1}{c}{Parameter} &  \multicolumn{1}{c}{P18+BSC}   & \multicolumn{1}{c}{P18+SH0ES+H0LICOW} & \multicolumn{1}{c}{P18lens+BSC+RSD} 
\\\hline
%$\Omega^{(0)}_{dm}h^2$ & $0.1188^{+0.0010}_{-0.0009}$ &  & 
%\\\hline
%$\Omega^{(0)}_{b}h^2$ & $0.02246^{+0.00017}_{-0.00019}$ & & 
%\\\hline
%$\tau$ & $0.0599^{+0.0064}_{-0.0130}$ &   & 
%\\\hline
$H_0$ & $67.86^{+0.64}_{-0.46}$ & $69.29\pm 0.61$ & $67.67\pm 0.62$
%\\\hline
%$n_s$ & $0.9697\pm 0.0038$ &  & 
\\\hline
$\sigma_8$ & $0.8090^{+0.0110}_{-0.0090}$ &  $0.8097\pm 0.0086$ &  $0.7994\pm 0.0084$
\\\hline
$\lambda$ & $0.40^{+0.20}_{-0.24}$ & $<0.163$ &  $0.54^{+0.24}_{-0.17}$
\\\hline
$\beta$ & $0.0198^{+0.0100}_{-0.0120}$ & $0.0240^{+0.0150}_{-0.0120}$  & $0.0167^{+0.0085}_{-0.0100}$ 
%\\\hline
%$\chi^2_{min,{\rm CDE}}-\chi^2_{min,\Lambda}$ & &  & 
\\\hline
$\ln\,B_{{\rm CDE},\Lambda}$ & $-5.88$ & $-6.54$ &  $-5.33$
\\\hline
 \end{tabular}}
\caption{\scriptsize{Mean values and $68\%$ c.l. uncertainties for the relevant parameters of the CDE model with exponential potential \eqref{eq:EXP}, obtained with three alternative data sets. See the corresponding comments in the main text of appendix C.}}
\label{tab:tab3}
\end{center}
\end{table}

%%%%%%%%%%%%%%%%%%%%%%%%%%%%%%%%%%%%%%%%%%%%%%%%%%%%%%%%%%%%%%%%%
%%%%%%%%%%%%%%%%%%%%%%%%%%%%%%%%%%%%%%%%%%%%%%%%%%%%%%%%%%%%%%%%%
%%%%%%%%%%%%%%%%%%%%%%%%%%%%%%%%%%%%%%%%%%%%%%%%%%%%%%%%%%%%%%%%%

\bibliographystyle{apsrev4-1}
\bibliography{UpdateCDE}

\end{document}